\documentstyle[sprocl]{article}

\def\Journal#1#2#3#4{{#1} {\bf #2}, #3 (#4)}

\def\NPB{{\em Nucl. Phys.} B}
\def\PLB{{\em Phys. Lett.}  B}

\def\PRD{{\em Phys. Rev.} D}

\def\be{\begin{equation}}
\def\ee{\end{equation}}
\def\bea{\begin{eqnarray}}
\def\eea{\end{eqnarray}}

\begin{document}

\title{
 NOVEL INSTABILITY IN SUPERSTRING COSMOLOGY
}

\author{JIRO SODA  and   MASA-AKI SAKAGAMI}

\address{Department of Fundamental Sciences, FIHS, Kyoto University
  \\ Kyoto 606-8501, Japan\\E-mail: jiro@phys.h.kyoto-u.ac.jp
   \\E-mail:sakagami@phys.h.kyoto-u.ac.jp} 

\author{SHINSUKE KAWAI}

\address{Graduate School of Human and Environmental Studies  \\ Kyoto
  University, Kyoto 606-8501, Japan\\E-mail:kawai@phys.h.kyoto-u.ac.jp}


\maketitle\abstracts{
 As a most promising candidate for quantum theory of the gravity, the
 superstring theory has attracted many researchers including
 cosmologists. It is expected that the cosmological initial
 singularity is avoided  within the context of the superstring
 theory. Indeed, Antoniadis et. al. found an interesting example of
 the the non-singular cosmological solutions by considering the
 string one-loop correction. Here, we will discuss the stability of
 this model and find a new kind of instability in the graviton modes. 
 We also argue this instability will persist even in the
 non-perturbative regime. The instability we have found might provide
 a mechanism to produce the primordial black holes.  
}

\section{Introduction}
As is well known, the singularity is unavoidable consequence of the
general relativity. The notorious cosmological singularity is nothing
but the manifestation of this general theorem. Usually, we expect that 
the quantum gravity works at the Planck scale and the singularity
disappear. To date, the superstring theory is the most promising
candidate for the consistent theory of quantum gravity. Hence, it is
interesting to seek the non-singular cosmological solutions in the low 
energy effective action of the superstring theory. From this point of
view, Antoniadis et. al.~\cite{anto} studied the one-loop corrected heterotic
superstring effective action and found a particular interesting class
of solutions which avoid the initial singularity. 

The purpose of this work is to show that these solutions are unstable
due to the high frequency gravitational wave. It means that the
background geometry is not viable in the most cases. 
 When we allow the fine tuning, however, the instability leads to an
 interesting consequence. 
If, at the arbitrary
initial time, the amplitude is sufficiently small, then the small
scale high amplitude gravitational waves will be produced by the
instability without destroying the background geometry. 
 In the deceleration phase, they will collapse and become the
primordial black holes. To support the above picture, we will discuss the 
non-perturbative effects phenomenologically and demonstrate that the
instability will persist even in the non-perturbative regime.

\section{Non-singular Cosmological Solution}

From the non-linear sigma model approach, we can calculate the $\beta$ 
function of the coupling function. The Weyl invariance requires the
vanishing of the $\beta$ function, which gives the equations of motion 
for the massless modes. From the equations of motion, the effective
action can be easily read off. When we fix the world surface as the
sphere, we obtain the tree level effective action. In actual
calculation, we usually expand the effective action with respect to
the string tension which controls the field theoretical loop
expansion. The string loop correction is different from this. The
string loop expansion corresponds to the path-integral quantization on 
the higher genus manifold. In particular, the one-loop string
correction can be obtained from the calculation of the $\beta$
function on the torus. It is this kind of correction that we want to
investigate in detail. 

The action for the heterotic superstring with orbifold
compactification is given by~\cite{anto}
\be
 S= \int \sqrt{-g}\{ {1\over 2}R- {1\over 4}(D\Phi )^2
    -{3\over 4}(D\sigma )^2 +{1\over 16}[\lambda_1 e^{\Phi}
    -\lambda_2 \xi (\sigma)] R_{GB}^2 \}
\ee
where $R,\Phi$ and $\sigma$ are the scalar curvature, dilaton, and the 
 modulus field, respectively. The Gauss-Bonnet curvature is given by 
\be
  R_{GB}^2 = R^{\mu\nu\rho\sigma} R_{\mu\nu\rho\sigma}
        -4 R^{\mu\nu} R_{\mu\nu} +R^2 
\ee
and the coupling function can be written by using the Dedekind $\eta$
function:
\be
  \xi (\sigma) = -\log [2e^{\sigma} \eta (ie^\sigma )] =\xi (-\sigma)
\ee
For the existence of the non-singular cosmological solutions,
the modulus field $\sigma$ is relevant. Therefore, from now on, we
will ignore the dilaton part of the action. Thus, we shall examine the 
modulus part of the effective action~\cite{anto,rizos}
\be
  S= \int d^4 x \sqrt{-g} \{
     {1\over 2}R- {1\over 4}(D\phi )^2
     -{1\over 16}
    \lambda \xi (\phi) R_{GB}^2 \}  \ .
\ee
Under the homogeneous and isotropic ansatz
\be
 ds^2 = -dt^2 + a^2 (t)\delta_{ij} dx^i dx^j   \ ,
\ee
we obtain the equations of motion:
\bea
 -3 H^2 &=& -{1\over 2} {\dot \phi}^2 
       -{3\lambda \over 2}H^3 {\dot \xi}  \\
 -[2{\dot H}+3H^2] &=& {1\over 2} {\dot \phi}^2 
       -{\lambda \over 2}[H^2 {\ddot \xi} + 2H{\dot \xi} 
         ({\dot H}+H^2) ] \ ,
\eea
where $ H= {\dot a}/a = da/dt/a $.

 It is convenient to define the effective energy density and pressure
 as $\rho_{\rm eff} = - G^0_0 $ and $p_{\rm eff} = G^0_0 /3$.
 Then,
\be
 \Gamma = {\rho_{\rm eff} +p_{\rm eff} \over \rho_{\rm eff}}
        = - {2{\dot H} \over 3 H^2}  \ .
\ee
Here, we would like to stress that this effective adiabatic index is
that of the background effective matter. It is not related to the
adiabatic index of the perturbative matter.  

Utilizing the asymptotic form
\be
  \xi_{,\phi} \sim {\rm sign}(\phi ) {\pi \over 3} e^{|\phi|} , 
\ee
it is possible to obtain the asymptotic solutions by imposing the
ansatz
\bea
  H&=& \omega_1 |t|^{\beta}  \\ 
  \phi &=& \phi_0 + \omega_2 \log |t|  \  .
\eea
In the asymptotic future, 
we expect the Friedman-Robertson-Walker universe, hence 
assume that the Einstein part balances with the ordinary energy 
 momentum in the  future. 
The asymptotic solution in $t \rightarrow \infty $ becomes 
\bea
  a &\sim & t^{1\over 3}  \ ,  \\ 
    H &=& {1\over 3}{1\over t} \ . 
\eea
This is the Friedman-Robertson-Walker solution with stiff matter. 
For the later purpose, we need to calculate the effective adiabatic
index.  
From the asymptotic solutions, it is easy to calculate $\Gamma$ as
\be
  \Gamma = 2     \ , t\rightarrow \infty    \ .
\ee

In the asymptotic past, $t\rightarrow -\infty $,  in order to obtain the
 non-singular solution, the Gauss-Bonnet part must balance with
 the ordinary energy momentum. 
 Then, we obtain the super-inflationary phase
\be
 H \sim {1\over (-t)^2 }  \ .
\ee
This type of solutions appears in the pre-big-bang scenario. The
serious problem there is the graceful exit problem. Interestingly,
one-loop corrected action gives a solution to this problem. 
The effective adiabatic index becomes 
\be
  \Gamma = {4t \over 3\omega_1 } \ , t\rightarrow -\infty \  .
\ee
Numerical calculation tells us that the above asymptotic solutions are 
smoothly connected. This is an example of the graceful exit in the
pre-big-bang scenario.~\cite{vene}

 Apparently, in the Gauss-Bonnet dominated phase,  
 the energy condition is violated
strongly. This fact itself is expected generally to obtain the non-singular
cosmological solutions.

\section{Stability analysis }

Intuitively, it is expected to have the instability in the background
which breaks the energy condition. In case of the perfect fluid,
the negative pressure causes the instability and the inhomogeneity is
enhanced catastrophically. However, what kind of instability can we
imagine in this field theoretical context? Does it happen in the
scalar perturbation, vector perturbation, or tensor perturbation?
Unfortunately, we do not have any answer without calculation. Indeed,
we performed numerical calculations and found no essential instability 
 in the scalar and vector perturbation.  It is the tensor perturbation 
 mode that shows the instability.
 Here is something interesting. The isotropic and
 homogeneous background drives the instability of the gravitational
 wave modes. This counter intuitive result prevents us to use the fluid 
 analogy. After all, the analogy is nothing but analogy. 
 Another interesting feature of this
 instability is its scale dependence. On the contrary to the Jeans
 instability,  the instability becomes strong in the high frequency
 graviton modes. Hence, the time scale of the instability could be
 arbitrarily small as far as the cut off scale is not introduced. 

To illustrate the instability explicitly, it is convenient to see the
Hamiltonian of the system. 
Consider the tensor perturbation
\be
  ds^2 = -dt^2 + a^2 (\delta_{ij} +h_{ij}) dx^i dx^j
\ee
where $h^{ij}_{,j} = h^i_i=0 $. By using the formula listed in the
Appendix,  we obtain
the action for the tensor perturbation as 
\bea
  S &=& {1\over 8} \int d^4 x a^3 [\dot{h}_{ij}\dot{h}^{ij}
      -{1\over a^2}\nabla {h_{ij}}\nabla {h^{ij}}
      -(4\dot{H}+6H^2 +\dot{\phi}^2 ){h_{ij}}{h^{ij}}] \nonumber \\
      & &  \qquad   -{\lambda \over 16}\int d^4 x a^3 \{ \ddot{\xi}
       [-a \nabla {h_{ij}}\nabla {h^{ij}} -2a^3 H^2 
          \dot{h}_{ij}\dot{h}^{ij} ]  \nonumber \\
      & &  \qquad -\dot{\xi} [-a^3 H\dot{h}_{ij}\dot{h}^{ij}
          +4a^3(\dot{H} +H^2) H {h_{ij}}{h^{ij}} ] \} \\
    &=& {1\over 8} \int d^4 x a^3 [
        (1-{\lambda \over 2}H\dot{\xi})\dot{h}_{ij}\dot{h}^{ij}
       -(1-{\lambda \over 2}\ddot{\xi}){1\over a^2}
        \nabla {h_{ij}}\nabla {h^{ij}} ]
\eea
The last expression is obatined by using the background eqations.
 From the background equation, we also get
\bea
  1-{\lambda \over 2}\ddot{\xi} 
    &=&(1-{\lambda \over 2}H\dot{\xi})(-{2\dot{H}\over H^2} -5)
    \nonumber \\
    &=&(1-{\lambda \over 2}H\dot{\xi})(3\Gamma-5)  
\eea
Notice that $\alpha \equiv 1-{\lambda \over 2}H\dot{\xi} >0$.
Finally, we have obtained the action for the graviton modes:
\be
  S= {1\over 8}\int d^4 x a^3 \alpha [\dot{h}_{ij}\dot{h}^{ij}
     -(3\Gamma -5) {1\over a^2} \nabla {h_{ij}}\nabla {h^{ij}} ]
\ee
 It is straightforward to deduce the Hamiltonian of the system in the
 following form:
\be
 H=\int d^4 x [ {2\pi^{ij} \pi_{ij} \over a^3 \alpha} 
      + (3\Gamma -5)a\alpha \nabla {h_{ij}}\nabla {h^{ij}} ]  \ ,
\ee
where $ \pi^{ij} = a^3 \alpha \dot{h}^{ij}/4$. 
 
If the energy condition is violated, i.e. $\Gamma <0 $ or $\Gamma
<{2\over 3} $, the system is unstable. Instability becomes strong as
the wavelength becomes short.

In case of one-loop model, we have $\Gamma =2$ in the
Friedmann-Robertson-Walker phase, so the background geometry
 is stable as is expected. On the other hand, it turns out that the
 Gauss-Bonnet phase is unstable due to the strong violation of 
the energy condition, $\Gamma \sim -|t|$. Strictly speaking, this
instability immediately does not imply the breakdown of the background
geometry. The point is that the fine tuning of the graviton amplitude
is required to avoid the breakdown of the background geometry.
 It is possible to give the criterion for the
breakdown of the background geometry explicitly.~\cite{kawai}

\section{Discussion}

First of all, it should be noted that the appearance of the 
Gauss-Bonnet term is generic features of the superstring cosmology.
 In case of the one-loop correction, the function form of the
 coupling between the modulus and the Gauss-Bonnet term is universal.
 Hence, our result can be regarded as the general feature at this level. 
 However,  our result suggest that 
 the non-perturbative effects should be taken
into account if superstring theory is viable for solving the
singularity problem. Assuming the Damour-Polyakov type universality,
 we can obtain the non-perturbative effective action
\be
  S = \int d^4 x \sqrt{-g} \{ R - (\partial \Phi )^2
        + B(\Phi) R_{GB}^2 + \cdots \}
\ee
where
\be
  B(\Phi) = e^{\Phi} +c_0 +c_1 e^{-\Phi} +c_2 e^{-2\Phi} + \cdots  \ .
\ee
Here, the function $B(\Phi )$ contains the correction from the string 
higher loops. 
The coefficients $c_i$ should take the values so that the initial
singularity will be avoided. The energy condition is necessarily
violated in this case again.  Now, we shall show, even in this case,
the instability we have found will appear. 
More generally, we can take into account other massless fields in the
model:
\be
  S = {1\over 2}\int d^4 x \sqrt{-g} [ R 
      - \sum_i (\nabla \phi_i )^2 -{\lambda \over 8}
       G (\phi_i ) R_{GB}^2 ]
\ee
where $G (\phi_i )$ is now a function of massless fields $\phi_i$.
The action for the tensor perturbation again takes the following form 
\be
 S_{\rm perturb} =
     {1\over 8}\int d^4 x a^3 (1-{\lambda \over 2}H \dot{G} (\phi_i  ))
        [\dot{h}_{ij}\dot{h}^{ij}
     -(3\Gamma -5) {1\over a^2} \nabla {h_{ij}}\nabla {h^{ij}} ] \ .
\ee
This is essentially the same as the previous form. 
 So, we have proved the existence of the instability. 

Next, we will argue that the fine tuning problem is not avoidable in
the non-singular cosmology in this model.  
Let us assume the non-singular cosmological solution. Then, we have
the eternal past and the eternal future.  Of course, there exists a
period where the energy condition is violated in order to have the
non-singular cosmological solution. However, 
 it is not allowed to have the infinite period of
violating the energy condition, otherwise we will encounter  the fine tuning
problem of the graviton amplitudes. 
 This implies that the effective adiabatic index $\Gamma $ is 
positive , more precisely $\Gamma >5/3$ in  most of the time. 
 From the definition 
$\Gamma = - {2\dot{H}/3H^2 }$, we obtain
\be
  {d \over dt}({1\over H(t)}) = {3\over 2}\Gamma (t) 
\ee
As $\Gamma (t) $ is almost always  positive, $1/H(t)$ is almost monotonic
function, hence $1/H(t)$ becomes zero at a certain time $t_0$. This
means 
\be
  \lim_{t\rightarrow t_0 } H(t) = \infty
\ee
namely we have encountered the singularity. Thus, as far as the
non-singularity and the avoidance of the fine tuning is assumed,  
 we must have the singularity somewhere. This is the contradiction.
Hence, if we want to have the non-singular cosmological solutions in
this context, the fine tuning problem can not be avoidable. 
Hence, this kind of non-perturbative correction can not change the 
situation.
It should be noted that the spatial curvature might change the
arguments above.~\cite{maeda}

An interesting  consequence of our result is that the instability is
stronger in the small scale.  
Hence, the instability found by us can provide a mechanism to generate the
primordial black holes. We will investigate this possibility in the
future.

\section*{Acknowledgments}
This work of M.S. is supported in part by Grant-in-Aid for Scientific
Research 09226220. 

\section*{Appendix}

Here, we will list the necessary formula to calculate the perturbative 
action:
\bea
  R^{(0)0j}_{0i} R^{(2)0i}_{0j} &=& (\dot{H} +H^2 )[
   -{1\over 2} h^{ik} \ddot{h}_{ik} -{1\over 4}\dot{h}^{ik}
     \dot{h}_{ik} -H h^{ik} \dot{h}_{ik} ] \nonumber  \\
  R^{(0)kl}_{ij} R^{(2)ij}_{kl} &=& 2 H^2 [
   -{1\over 4 a^2} h^{ik,m} h_{ik,m} -{1\over 4}\dot{h}^{ik}
     \dot{h}_{ik} -2H h^{ik} \dot{h}_{ik} ] \nonumber \\
  R^{(1)0j}_{0i} R^{(1)0i}_{0j} &=&
   {1\over 4} \ddot{h}^{ik} \ddot{h}_{ik} +H \dot{h}^{ik}
     \ddot{h}_{ik} +H^2 \dot{h}^{ik} \dot{h}_{ik} \nonumber \\
  R^{(1)0j}_{kl} R^{(1)kl}_{0j} &=&
   -{1\over 2 a^2} \dot{h}^{ik,m} \dot{h}_{ik,m}  \nonumber \\
 R^{(1)ij}_{kl} R^{(1)kl}_{ij} &=&
   {1\over a^4} \nabla^2 h^{ik} \nabla^2 h_{ik} -{2H\over a^2}\dot{h}^{ik}
     \nabla^2 h_{ik} +H^2 \dot{h}^{ik} \dot{h}_{ik} \nonumber \\
 R^{(0)0}_{0} R^{(2)0}_{0} &=&
     3(\dot{H} +H^2 )[
   -{1\over 4} \dot{h}^{ik} \dot{h}_{ik} -{1\over 2} h^{ik}
     \ddot{h}_{ik} -H h^{ik} \dot{h}_{ik} ]  \nonumber \\
 R^{(0)i}_{j} R^{(2)j}_{i} &=&
     (\dot{H} +3H^2 )[
   -{1\over 2} h^{ik} \ddot{h}_{ik} -{1\over 2} \dot{h}^{ik}
     \dot{h}_{ik} -3H h^{ik} \dot{h}_{ik} 
      -{1\over 4a^2} h_{jk,m} h^{jk,m}]   \nonumber \\
 R^{(1)i}_{j} R^{(1)j}_{i} &=&
   {1\over 4} \ddot{h}^{ik} \ddot{h}_{ik} 
     +{1\over 4a^4} \nabla^2 h_{jk} \nabla^2 h^{jk}
   +{9\over 4}H^2  \dot{h}^{ik}  \dot{h}_{ik}  \nonumber \\ 
   & & \qquad -{1\over 2a^2} \ddot{h}_{jk} \nabla^2 h^{jk} 
   +{3\over 2}H \ddot{h}^{ik} \dot{h}_{ik} 
   -{3H \over 2a^2}  \dot{h}_{jk} \nabla^2 h^{jk}]  \nonumber \\
 R^{(0)} R^{(2)} &=&
    6(\dot{H} +2H^2 )[
   -{3\over 4} \dot{h}^{ik} \dot{h}_{ik} - h^{ik}
     \ddot{h}_{ik} -4H h^{ik} \dot{h}_{ik} 
      -{1\over 4a^2} h_{jk,m} h^{jk,m}]   \ . \nonumber  
\eea
Here, we have omitted the spatially total derivative terms, because
they do not contribute the final answer.

\end{document}